\documentclass[twocolumn,aps,superscriptaddress, showkeys, nofootinbib,floatfix,linenumbers]{revtex4}
\usepackage{epsfig,bm,feynmf}
\usepackage{graphics}
\usepackage{amsmath}
\usepackage{mathrsfs}
\usepackage{appendix}
\usepackage{bm}
\usepackage[normalem]{ulem}  
\usepackage[dvips]{color} 
\newcommand{\com}[1]{{\sf\color[rgb]{0,0,1}{#1}}}

\renewcommand{\sout}{\bgroup \color{red} \ULdepth=-.5ex \ULset}

\begin{document}
\title{Lambda hyperon polarization in relativistic heavy ion collisions from the chiral kinetic approach}

\author{Yifeng Sun}
\email{sunyfphy@physics.tamu.edu}
\affiliation{Cyclotron Institute and Department of Physics and Astronomy, Texas A$\&$M University, College Station, Texas 77843, USA}%

\author{Che Ming Ko}
\email{ko@comp.tamu.edu}
\affiliation{Cyclotron Institute and Department of Physics and Astronomy, Texas A$\&$M University, College Station, Texas 77843, USA}%

\date{\today}

\begin{abstract}
Based on the chiral kinetic approach using initial conditions from a multiphase transport model, we study the spin polarizations of quarks and antiquarks in non-central heavy ion collisions at the Relativistic Heavy Ion Collider.  Because of  the non-vanishing vorticity field  in these collisions, quarks and antiquarks are found to acquire appreciable spin polarizations in the direction perpendicular to the reaction plane.  Converting quarks and antiquarks to hadrons via the coalescence model, we further calculate the spin polarizations of Lambda and anti-Lambda hyperons and find their values comparable to those measured in experiments by the STAR Collaboration.
\end{abstract}

\maketitle

\section{introduction}
Because of the large orbital angular momentum in non-central collisions of two heavy nuclei at relativistic energies, the vorticity field in the produced quark-gluon plasma (QGP) can reach a very large value of $10^{21}$ s$^{-1}$~\cite{PhysRevC.93.064907,PhysRevC.94.044910}. It has been argued that due to their spin-orbit interactions, quarks and antiquarks can be polarized along the direction of the orbital angular momentum by the vorticity field, and after hadronization, they then lead to the production of polarized Lambda ($\Lambda$) hyperons, which can be measured in experiments~\cite{PhysRevLett.96.039901,PhysRevC.77.044902}.   Taking into consideration of the non-uniformity of the vorticity field through that in the spatial distribution of particles in the statistical-hydrodynamic approach~\cite{PhysRevC.77.024906,Becattini20082452,Becattini201332} or the quantum kinetic approach~\cite{PhysRevC.94.024904}, local structures in the polarization density can also be studied.  Such effects have been included in many studies based on the hydrodynamical model~\cite{PhysRevC.93.064907,PhysRevC.94.044910,PhysRevC.88.034905,Karpenko2017,PhysRevC.94.054907,PhysRevC.95.031901,PhysRevC.87.034906,PhysRevC.90.021904,Becattini2015,PhysRevLett.117.192301,Aristova2016,PhysRevC.88.061901,PhysRevC.92.014906,Ivanov:2017dff}, and some of these studies can indeed quantitatively reproduce the polarization of $\Lambda$ hyperons measured by the STAR Collaboration~\cite{STAR:2017ckg}. A similar study using the transport model~\cite{Li:2017slc} also gives a result that agrees with the experimental observation.

Although the time evolution of the vorticity field is included in both the hydrodynamic and transport studies, the spin polarization of hyperons is calculated from their equilibrium distribution in the final state and is therefore only determined by the final vorticity field in the produced matter. Moreover, these studies have not addressed the non-equilibrium effect on the polarization of $\Lambda$ hyperons.   To include the non-equilibrium effect on the spin polarizations of quarks and antiquarks in the partonic phase of heavy ion collisions, we use in the present study the chiral kinetic approach that takes into account the effect of the vorticity field on the equations of motion~\cite{PhysRevLett.109.162001,PhysRevLett.109.232301,PhysRevLett.110.262301,PhysRevLett.109.181602,PhysRevD.89.094003,PhysRevD.90.076007,PhysRevLett.113.182302,PhysRevLett.115.021601} and scatterings of quarks and antiquarks.  By converting the resulting polarized quarks and antiquarks to hadrons using the coalescence model after the partonic phase~\cite{PhysRevLett.90.202302,Ko2014234}, we further study the polarizations of $\Lambda$ and $\bar\Lambda$ hyperons.

This paper is organized as follows. In the next section, we describe the chiral kinetic equation of motion for quarks and antiquarks and their modified scattering in the presence of a vorticity field. We then introduce in Sec.III the coalescence model that takes into account explicitly the spin degrees of freedom in converting polarized quarks and anitiquarks to polarized $\Lambda$ and $\bar\Lambda$ hyperons.  In Sec. IV, we describe how a multiphase transport model (AMPT)~\cite{PhysRevC.72.064901} is used to generate the initial phase-space distributions of quarks and antiquarks and the way their vorticity field is calculated.  We show in Sec. V results on the time evolution of the spin polarizations of quarks and antiquarks as well as those of $\Lambda$ and $\bar\Lambda$ hyperons.  Results on the rapidity, transverse momentum, and collision energy dependence of the $\Lambda$ and $\bar\Lambda$ polarizations are then shown and compared with experimental data.  Finally, a summary is given in Sec. VI.

\section{Chiral kinetic approach}

In this section, we introduce the chiral kinetic equations of motion for massless spin-$\frac{1}{2}$ particles and their modified scatterings in a vorticity field that are used in the present study of spin polarizations in heavy ion collisions.

\subsection{The chiral kinetic equation}

As pointed out in Ref.~\cite{PhysRevLett.109.162001}, the effect of the vorticity field on the motions of massless fermions can be included in the chiral kinetic equation through the Coriolis force. In the absence of mean-field potentials, such as those in Refs.~\cite{Ko2014234,Xu:2013sta,Xu:2014qra,Xu:2016ihu} based on the Nambu-Jona-Lasinio model, a particle of momentum ${\bf p}$ in the center-of-mass or fireball frame of a heavy ion collision is a constant between scatterings, that is $\dot{\bf p}=0$.  Its momentum ${\bf p}^\prime$ in the frame that rotates with an angular velocity given by the vorticity field $\boldsymbol\omega$ is, however, affected by the Coriolis force,
\begin{eqnarray}\label{coriolis}
\dot{\bf p}^\prime=-2p^\prime\boldsymbol{\omega}\times\dot{\bf r}^\prime,
\end{eqnarray}
where $p^\prime$ is the magnitude of ${\bf p}^\prime$ and $\dot{\bf r}^\prime$ is the velocity of the particle in the rotating frame.

From the adiabatic approximation to the motion of a massless spin-$\frac{1}{2}$ particle that the direction of its spin follows instantaneously the direction of its momentum~\cite{PhysRevLett.109.162001} or from considering the conservation of the total angular momentum of the particle~\cite{PhysRevC.94.045204} in this rotating frame, its velocity $\dot{\bf r}^\prime$ is given by,
\begin{eqnarray}\label{chiral}
\dot{\bf r}^\prime=\hat{\bf p}^\prime+\lambda\dot{\bf p}^\prime\times{\bf b}^\prime,
\end{eqnarray}
where $\lambda=\pm 1$ is the helicity of the particle, $\hat{\bf p}^\prime$ is a unit vector along ${\bf p}^\prime$, and ${\bf b}^\prime=\frac{\hat{\bf p}^\prime}{2{p^\prime}^2}$ is the Berry curvature resulting from the adiabatic approximation~\cite{PhysRevLett.109.162001} or the conservation of the total angular momentum~\cite{PhysRevC.94.045204}.  Substituting Eq.(\ref{coriolis}) to Eq.(\ref{chiral}) leads to
\begin{eqnarray}\label{rotating}
\dot{\bf r}^\prime&=&\frac{\hat{\bf p}^\prime+2\lambda p^\prime(\hat{\bf p}^\prime\cdot{\bf b}^\prime)\boldsymbol\omega}{1+2\lambda p^\prime(\boldsymbol\omega\cdot{\bf b}^\prime)}.
\end{eqnarray}

In the absence of collective flow, the origin of the rotational frame is stationery in the fireball frame. In this case, $\dot{\bf r}=\dot{\bf r}^\prime$ and ${\bf p}={\bf p}^\prime$, and the chiral kinetic equations of motion for massless spin-$\frac{1}{2}$ particles are thus
\begin{eqnarray}
\dot{\mathbf{r}}=\frac{\hat{\mathbf{p}}+2\lambda p(\mathbf{\hat p}\cdot{\mathbf b})\boldsymbol{\omega}}{1+2\lambda p(\boldsymbol{\omega}\cdot{\mathbf b})},\quad\dot{\bf p}=0.
\label{CKE}
\end{eqnarray}

The two equations in Eq.(\ref{CKE}) are similar to those derived in Ref.~\cite{PhysRevLett.110.262301} from consideration of the covariant Wigner distribution function of massless spin 1/2 particles in a vorticity field, except that the numerical factor 2 in the denominator is 4 instead. Although applying the usual Lorentz transformation to Eq.(\ref{rotating}) does not lead to Eq.(\ref{CKE}), it is known that the chiral kinetic equations satisfy a nontrivial Lorentz transformation~\cite{Chen:2014cla}.  Therefore we assume that Eq.(\ref{CKE}) is also valid for the case with collective flow.  As to the numerical factor in the denominator, later considerations based on the covariant Wigner distribution function of massless spin 1/2 particles have indicated that its value  is not uniquely determined~\cite{Gao:2017gfq}. To decide on this numerical factor, we consider in the following the effect of the Berry curvature on the measure of phase-space.

\subsection{Vorticity-field modified collisions}

Besides modifying the equations of motion for massless fermions, the vorticity field also affects their distributions in phase-space by introducing the measure $\sqrt{G}=1+a\lambda |\mathbf{p}|(\boldsymbol{\omega}\cdot\mathbf{b})$~\cite{PhysRevLett.109.162001,PhysRevLett.109.232301,PhysRevLett.110.262301,PhysRevLett.109.181602}, which  appears in the denominator of Eq.(\ref{CKE}) if $a=2$.  Taking into account the modified phase-space measure changes the particle density in a fluid with non-zero vorticity field from $\int \frac{d^3{\bf p}}{(2\pi)^3}f(p^0)$ to $\int \frac{d^3{\bf p}}{(2\pi)^3}\sqrt{G}f(p^0)$~\cite{Gao2015542}, where $p_0=\gamma(|\mathbf{p}|-\mathbf{v}\cdot\mathbf{p})$ with $\gamma=\frac{1}{\sqrt{1-\mathbf{v}^2}}$ and $\mathbf{v}$ being the fluid velocity. In the local rest frame of the fluid and assuming an equilibrium Boltzmann distribution at temperature $T$, the average spin of massless spin-$\frac{1}{2}$ fermions is then
\begin{eqnarray}
\mathbf{S}=\frac{\int\frac{d^3{\bf p}}{(2\pi)^3}\frac{\lambda}{2}\hat{\mathbf{p}}\sqrt{G}
f(p)}{\int\frac{d^3{\bf p}}{(2\pi)^3}\sqrt{G}f(p)}=\frac{a}{6}\left(\frac{\lambda^2\boldsymbol{\omega}}{4T}\right),
\end{eqnarray}
which differs from the usual value by the factor $a/6$ as a result of the modified phase-space measure from the Berry curvature. This result shows that the vorticity field can lead to the spin polarization of a massless fermion along its direction, and the result agrees with that from the quantum kinetic approach~\cite{PhysRevC.94.024904} if the value of $a$ is taken to be $a=6$. In the present study, we thus use Eq.(\ref{CKE}) with the numerical factor 2 in the denominator replaced by the numerical factor 6 to describe the motions of massless fermions in a vorticity field.

The above equations of motion and modified phase-space measure are for massless fermions.  Since treating helicity as a good quantum number remains a good approximation for quarks of finite but small masses, their equations of motion can still be given by Eq.(\ref{CKE}) after replacing $\hat{\mathbf{p}}$, $p$, and ${\bf b}=\frac{\hat{\bf p}}{2p^2}$ by $\frac{\mathbf{p}}{E_p}$, $E_p$ and ${\bf b}=\frac{\hat{\bf p}}{2E_p^2}$, respectively~\cite{PhysRevD.89.094003}.

To ensure that massless fermions reach the equilibrium distribution $\sqrt{G}f((p^0-\mu)/T)$, where $\mu$ is the baryon chemical potential, from their collisions, the momenta $\mathbf{p_3}$ and $\mathbf{p_4}$ of two colliding fermions after a collision are determined by momentum conservation but with the probability $\sqrt{G(\mathbf{p_3})G(\mathbf{p_4})}$. This effect can be included in the calculation by randomly selecting their momenta after a collision in their center of mass frame according to momentum conservation and then Lorentz transforming to the fireball frame. The probability $\sqrt{G(\mathbf{p_3})G(\mathbf{p_4})}$ is then calculated and compared with a random number. This process is continued until the random number is less than $\sqrt{G(\mathbf{p_3})G(\mathbf{p_4})}$. Since the values of $\sqrt{G(\mathbf{p_3})}$ and $\sqrt{G(\mathbf{p_4})}$ can be negative, which is unphysical as a result of the approximation used in deriving the chiral vortical equation of motion, we set their values to one and treat these particles without the effect due to the vorticity field.  Also, the value of $\sqrt{G(\mathbf{p_3})G(\mathbf{p_4})}$ can be great than one. In this case, the random number mentioned above is taken between $0$ and $25$. Because of the upper cutoff, particles of small momentum are not affected by the modified phase-space measure.  Since the number of such particles is small, we have checked that increasing the value of the upper cutoff does not affect our results.  As to the conditions for the collision, they can be determined by using the geometric method based on the scattering cross section as described in Ref.~\cite{Bertsch:1988ik}.  For the collision between a fermion and its antiparticle that have opposite helicities, their helicities can be flipped after the collision. We include this chirality changing scattering in the present study as in our previous studies~\cite{PhysRevC.94.045204,PhysRevC.95.034909}.

\section{Coalescence model for production of polarized $\Lambda$ ($\bar\Lambda$) hyperon}

To study the spin polarizationof $\Lambda$ ($\overline{\Lambda}$) hyperon, we use the coalescence model for hadron production~\cite{PhysRevLett.90.202302} to convert polarized quarks and antiquarks to polarized baryons and antibaryons. In this model, the probability for $u$, $d$ and $s$ quarks (antiquarks) at phase-space points (${\bf r}_1$, ${\bf p}_1$), (${\bf r}_2$, ${\bf p}_2$), and (${\bf r}_3$, ${\bf p}_3$), respectively, to form a $\Lambda$ ($\overline{\Lambda}$) is given by the quark Wigner function of the $\Lambda$ ($\overline{\Lambda}$)~\cite{Ko2014234}:
\begin{eqnarray}\label{wigner}
&&f_\Lambda(\boldsymbol{\rho},\boldsymbol{\lambda},\mathbf{k_{\rho}},\mathbf{k_{\lambda}})=8^2g_C g_Se^{-\frac{\boldsymbol{\rho}^2}{\sigma_{\rho}^2}-\frac{\boldsymbol{\lambda}^2}{\sigma_{\lambda}^2}-\mathbf{k_{\rho}}^2\sigma_{\rho}^2-\mathbf{k_{\lambda}}^2\sigma_{\lambda}^2}\com{,}\sout{.}
\end{eqnarray}
if the wave functions of the quarks (antiquarks) are taken to be those of a harmonic oscillator potential. In the above, $g_C=1/27$ is the coalescence factor for colored $u$, $d$, and $s$ quarks to form a colorless $\Lambda$, $g_S$ is the coalescence factor for these polarized quarks to form a polarized $\Lambda$ and will be discussed later.  The relative coordinates and momenta are defined as
\begin{eqnarray}
\boldsymbol{\rho}&=&\frac{1}{\sqrt{2}}(\mathbf{r}^\prime_1-\mathbf{r}^\prime_2), \quad\mathbf{k_{\rho}}=\sqrt{2}\frac{m_2\mathbf{p}^\prime_1-m_1\mathbf{p}^\prime_2}{m_1+m_2},\nonumber\\
\boldsymbol{\lambda}&=&\sqrt{\frac{2}{3}}\left(\frac{m_1\mathbf{r}^\prime_1+m_2\mathbf{r}^\prime_2}{m_1+m_2}-\mathbf{r}^\prime_3\right),\nonumber\\
\mathbf{k_{\lambda}}&=&\sqrt{\frac{3}{2}}\frac{m_3(\mathbf{p}^\prime_1+\mathbf{p}^\prime_2)-(m_1+m_2)\mathbf{p}^\prime_3}{m_1+m_2+m_3},\\
&&\nonumber
\end{eqnarray}
where $m_i$ is the mass of $i$-th quark, and ${\bf r}^\prime_i$ and ${\bf p}^\prime_i$ are its position and momentum in the center-of-mass frame of the three quarks after they are propagated to the time when the last quark freezes out. The width parameters $\sigma_\rho$ and $\sigma_\lambda$ in Eq.(\ref{wigner}) are related to the oscillator frequency $\omega$ by $\sigma_{\rho}=1/\sqrt{\mu_1\omega}$ and $\sigma_{\lambda}=1/\sqrt{\mu_2\omega}$, respectively, where the reduce masses are $\mu_1=\frac{2}{1/m_1+1/m_2}$ and $\mu_2=\frac{3/2}{1/(m_1+m_2)+1/m_3}$.

It can be shown that the oscillator frequency $\omega$ is related to the matter radius $\langle r^2\rangle_\Lambda$ of $\Lambda$~\cite{Ko2014234},
\begin{eqnarray}
\langle r^2\rangle_\Lambda=\frac{1}{2\omega}
\frac{\sum_{(i,j)=(1,2),(2,3),(3,1)}m_im_j(m_i+m_j)}{m_1m_2m_3(m_1+m_2+m_3)}.
\end{eqnarray}
Taking the root-mean-squared radius of $\Lambda$ to have a value of 0.877 fm, similar to that of a proton as in Ref.~\cite{Ko2014234}, leads to $\sigma_{\rho}= 0.89$ fm and $\sigma_{\lambda}=0.51$ fm if the masses of $u$, $d$ and $s$ quarks and antiquarks are taken approximately to be 3, 6, and 100 MeV, respectively~\cite{Olive:2016xmw}.

To evaluate $g_S$ in Eq.(\ref{wigner}), we note that the spin state of $\Lambda$ is determined by the spin state of its constituent $s$ quark, so its value is given by the probability for $u$ and $d$ quarks to form a spin-singlet state.  Denoting the spin vectors of $u$ and $d$ quarks in the fireball frame by $\lambda_1\hat{\mathbf{p}}_1/2$ and $\lambda_2\hat{\mathbf{p}}_2/2$, respectively, where their respective momenta are $\mathbf{p}_1$ and $\mathbf{p}_2$ and respective helicities are $\lambda_1$ and $\lambda_2$, the probability for $u$ and $d$ quarks to form a spin-singlet state is then
\begin{eqnarray}
g_S&=&|(1/\sqrt{2})(\langle\uparrow\downarrow|-\langle\downarrow\uparrow|)[\cos(\theta_1/2)\cos(\theta_2/2)|\uparrow\uparrow\rangle\nonumber\\
&&+\cos(\theta_1/2)\sin(\theta_2/2)e^{i\phi_2}|\uparrow\downarrow\rangle\nonumber\\
&&+\sin(\theta_1/2)\cos(\theta_2/2)e^{i\phi_1}|\downarrow\uparrow\rangle\nonumber\\
&&+\sin(\theta_1/2)\sin(\theta_2/2)e^{i(\phi_1+\phi_2)}]|\downarrow\downarrow\rangle|^2\nonumber\\
&=&\frac{1}{4}(1-\cos\theta_1\cos\theta_2-\sin\theta_1\sin\theta_2\cos(\phi_1-\phi_2))\nonumber\\
&=&\frac{1}{4}(1-\lambda_1\lambda_2\hat{\mathbf{p}}_1\cdot\hat{\mathbf{p}}_2),
\end{eqnarray}
where $\theta_1$ ($\theta_2$) and $\phi_1$ ($\phi_2$) are the azimuthal angles of the spin vector $\lambda_1\hat{\mathbf{p}}_1/2$ ($\lambda_2\hat{\mathbf{p}}_2/2$) of $u$ ($d$) quark.

\section{Initial conditions and the vorticity field}

To study the effect of the vorticity field in relativistic heavy ion collisions based on the chiral kinetic approach, we use the string-melting version of AMPT model~\cite{PhysRevC.72.064901}, with its parameters taken from Ref.~\cite{PhysRevC.84.014903}, to generate the initial phase-space distributions of quarks and antiquarks. Specifically, we use the values $a=0.5$ and $b=0.9$~GeV$^{-2}$ in the Lund string fragmentation function. We further use the impact parameter $b=8.87$ fm to correspond to the centrality bin of $20-50\%$ in Au+Au collisions, which is obtained from the geometrical relation $c=\pi b^2/\sigma_{\rm in}$ with $\sigma_{\rm in}=705$ fm$^2$ between the centrality bin and the impact parameter~\cite{PhysRevC.65.024905}. For the helicities of initial partons, they are taken to have random positive or negative values, corresponding to an initial state of vanishing axial charge density and spin polarization.  The phase-space distributions of these quarks and antiquarks are then evolved in time according to the chiral equations of motion [Eq.(\ref{CKE})] and by scatterings with an isotropic and constant cross section of 10 mb before including the  modification by the vorticity field described above until they stop scattering.

The vorticity field $\boldsymbol{\omega}$ in Eq.(\ref{CKE}) is related to the velocity field $\mathbf{v}({\bf r},t)$ of partons by $\boldsymbol{\omega}=\frac{1}{2}\boldsymbol{\nabla}\times\mathbf{u}$ with $\mathbf{u}=\gamma\mathbf{v}$ and $\gamma=\frac{1}{\sqrt{1-\mathbf{v}^2}}$. For the velocity field, it can be calculated from the average velocity of partons in a local cell, i.e., $\mathbf{v}(\mathbf{r},t)=\frac{\sum_i \mathbf{p_i}/E_i}{\sum_i1}$ with $i$ running over all partons in this cell, similar to the velocity of the particle flow in Ref.~\cite{PhysRevC.93.064907}.   In our calculations, we divide the whole volume of the colliding system into $61\times61\times61$ cells with spacing $\Delta x=\Delta y=\Delta z=0.5$ fm.  Although partons are produced with different formation times, they are all included in calculating the local velocity field.

\section{Results}

In this section, we first solve the chiral kinetic equations for quarks and antiquarks from the AMPT model to obtain their average spin, $\mathbf{S}=\frac{\sum_i \frac{\lambda_i\hat{\mathbf{p}}_i}{2}}{\sum_i 1}$, and polarization along the $y$ direction, $P_y=2S_y$. We then use the coalescence model to convert these partons to $\Lambda$ and $\bar\Lambda$ hyperons and study their polarizations.

\subsection{Polarization of light quarks and antiquarks}

\begin{figure}[h]
\centering
\includegraphics[width=0.5\textwidth]{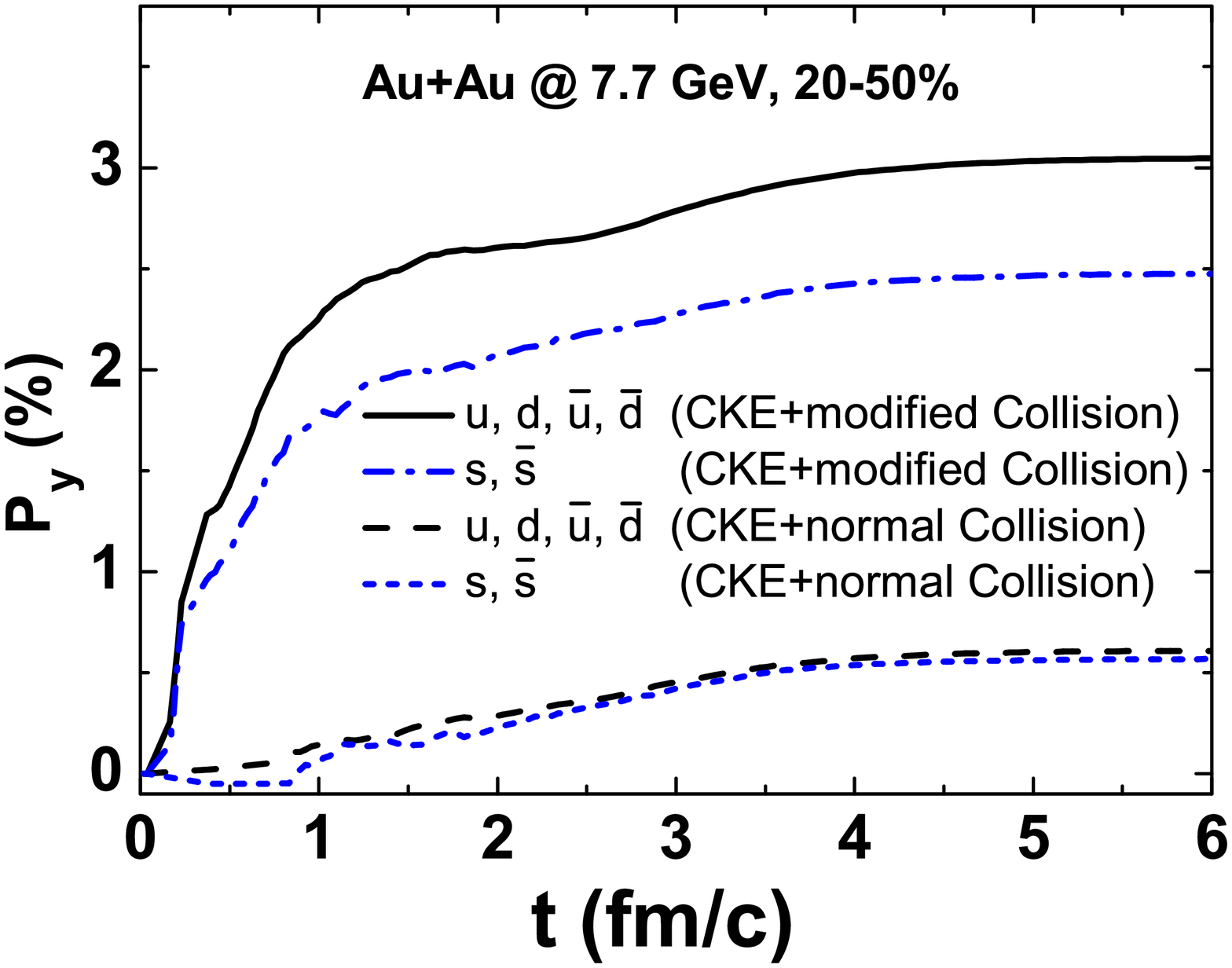}
\caption{Time evolution of the spin polarization of light and strange quarks and antiquarks in midrapidity $|y|\le1$ along the total orbital angular momentum with (solid lines) and without (dashed lines) using parton scattering cross section that includes the effect of local vorticity field.}
\label{pt}
\end{figure}

We first consider Au+Au collisions at $\sqrt{s_{NN}}=7.7$ GeV, which is the lowest energy in the beam energy scan (BES) program of the STAR Collaboration at RHIC.  Shown in Fig.\ref{pt} is the time evolution of the spin polarization along $y$ direction $P_y$ of light and strange quarks and antiquarks in midrapidity $|y|\le 1$.  Solid and dash-dotted lines are the respective results for light and strange quarks and antiquarks from using the chiral kinetic equations and the scattering cross section that includes the effect due to the modified phase-space measure. We see that the spin polarization of strange quarks and antiquarks is smaller than that of light quarks and antiquarks, and this is due to the larger strange quark mass than light quark masses and their different temporal and spatial distributions.  As to their time dependence, the spin polarizations of both strange and non-strange quarks and antiquarks increase rapidly from zero during the first 1 fm$/c$ and then gradually reach their respective constant values during $1-3.5$ fm/$c$. The initial rapid increase in the spin polarization is due to the fast approach of partons to their equilibrium distributions as a result of scatterings, while the slow increase in later times is because it takes a longer time to generate an axial charge dipole moment in the transverse plane from chiral kinetic motions, which is responsible for the generation of spin polarization~\cite{PhysRevC.95.034909}. This can be seen from the long- and short-dashed lines that the spin polarization increases much slower from zero to the maximum value if we use an isotropic cross section instead of the non-isotropic one because of the modified phase-space measure due to the  vorticity field.

\subsection{Polarization of $\Lambda$ and $\overline{\Lambda}$ hyperons }

\begin{figure}[h]
\centering
\includegraphics[width=0.5\textwidth]{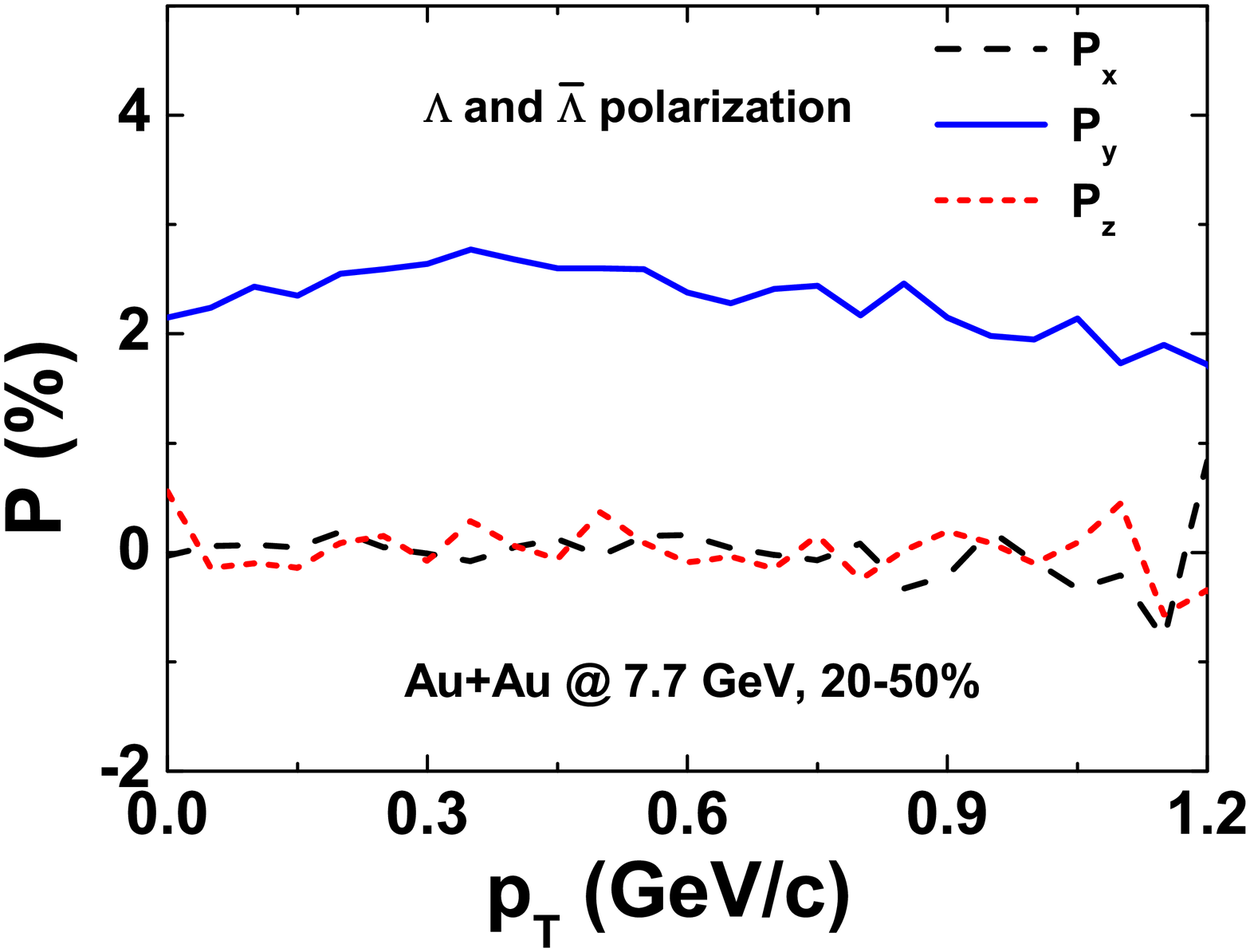}
\caption{Transverse momentum dependence of the spin polarization of midrapidity ($|y|\le 1$) $\Lambda$ and $\bar\Lambda$ hyperons along different directions in Au+Au collisions at $\sqrt{s_{NN}}=7.7$ GeV.}
\label{pts}
\end{figure}

Shown in Fig. \ref{pts} is the transverse momentum dependence of the spin polarization of $\Lambda$ and $\bar\Lambda$ hyperons at midrapidity ($|y|\le 1$) in different directions at the collision energy $\sqrt{s_{NN}}=7.7$ GeV. It is seen that its three components all shows little dependence on transverse momentum.  Although the values of $x$- and $z$-components are consistent with zero, the total spin polarization in the $y$-direction, i.e., along the direction of total orbital angular momentum, has a value of 2.44$\%$, which is almost the same as the spin polarization of strange quarks and antiquarks shown in Fig.~\ref{pt}.

\begin{figure}[h]
\centering
\includegraphics[width=0.5\textwidth]{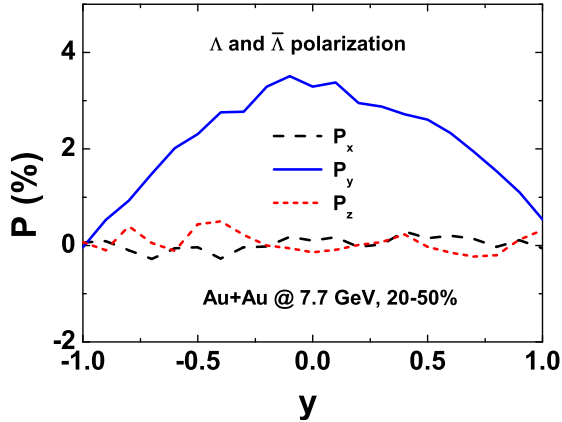}
\caption{Rapidity dependence of the spin polarization of midrapidity ($|y|\le 1$) $\Lambda$ and $\bar\Lambda$ hyperons along different directions in Au+Au collisions at $\sqrt{s_{NN}}=7.7$ GeV.}
\label{ys}
\end{figure}

The rapidity dependence of the spin polarization for $\Lambda$ and $\overline{\Lambda}$ hyperons is shown in Fig. \ref{ys} at the same collision energy. Although the spin polarizations along the $x$-direction and the $z$-direction are consistent with zero for all rapidities, that along the direction of total orbital angular momentum peaks at zero rapidity and decreases with increasing magnitude of rapidity. The latter is due to the smaller number of scatterings for quarks and antiquarks at large rapidity.

\begin{figure}[h]
\centering
\includegraphics[width=0.5\textwidth]{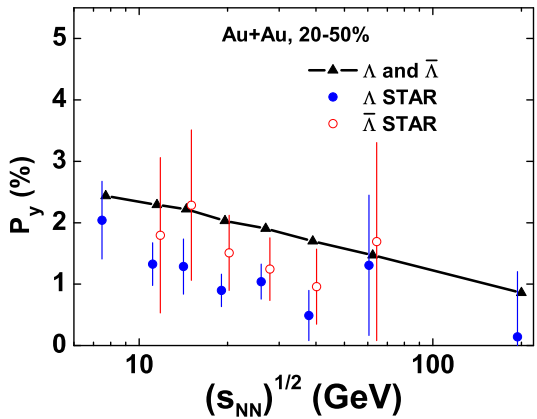}
\caption{Energy dependence of the spin polarization of midrapidity ($|y|\le 1$) $\Lambda$ and $\overline{\Lambda}$ hyperons in Au+Au collisions at energies from 7.7 GeV to 200 GeV. Data with error bars are from the STAR Collaboration~\cite{PhysRevC.76.024915,STAR:2017ckg}.}
\label{sp}
\end{figure}

In Fig.~\ref{sp}, we show the collision energy dependence of the spin polarization of $\Lambda$ and $\bar\Lambda$ hyperons along the direction of total orbital angular momentum. Our results indicate that it decreases with increasing collision energy, and this is due to the decrease of the vorticity field in the partonic matter at freeze-out as a result of both the decrease of the initial vorticity field and the longer lifetime of partonic phase, which leads to a decrease of the final vorticity field, with increasing collision energy~\cite{PhysRevC.93.064907,PhysRevC.94.044910,Karpenko2017}.  These results are similar in both trend and value to the experimental data~\cite{STAR:2017ckg} and results from other studies based on the statistical model after a hydrodynamic or transport evolution of the produced matter~\cite{Karpenko2017,Li:2017slc}.  There are also feed-down corrections from resonance decays, which are not included in the present study.  According to Refs.~\cite{Karpenko2017,Qiu:2013wca,Becattini:2016gvu}, including $\Lambda$ and $\bar\Lambda$ from resonance decays reduces the polarization of $\Lambda$ and $\bar\Lambda$ hyperons by 15\% to 20\%. On can see from Fig.~\ref{sp} that including this effect is expected to bring our results into better agreement with the data.

\section{Summary}

We have used the chiral kinetic equations to study the effect of vorticity field on the spin polarizations of light and strange quarks and antiquarks based on initial conditions from the AMPT model. We find that they all acquire non-vanishing spin polarizations as a result of the chiral vortical effects on their equations of motion. Their polarizations are further increased by the modified collisions due to the modification of their phase-space distributions.  Using the coalescence model to convert light and strange quarks and antiquarks to $\Lambda$ and $\bar\Lambda$ hyperons after the partonic phase, we have further found that their spin polarizations, which are determined by those of strange quarks and antiquarks, are comparable to the experimental data. However, we have not included the effect of hadronic evolution on the $\Lambda$ and $\bar\Lambda$ polarizations.  To study this effect requires a hadronic transport model that includes explicitly their spin degrees of freedom, such as in Refs.~\cite{Xu:2012hh,Xia:2014qva,Xia:2014rua}, and is beyond the present study.  Also, $\Lambda$ and $\bar\Lambda$ have same polarization in our study because they are described by the same chiral kinetic equations of motion and modified collisions.  The present study thus cannot explain the seemingly larger $\bar\Lambda$ than $\Lambda$ polarization seen in the experimental data. Furthermore,  we have only considered the global spin polarizations of  $\Lambda$ and $\bar\Lambda$ hyperons.  Since the transport model includes non-equilibrium effects and the local structure of vorticity field, it provides the possibility to study the local structure of the spin polarization~\cite{PhysRevLett.117.192301} and other interesting features of this phenomenon.

\section*{ACKNOWLEDGEMENTS}

This work was supported in part by the US Department of Energy under Contract No. DE-SC0015266 and the Welch Foundation under Grant No. A-1358.

\bibliography{ref}

\end{document}